\documentclass[12pt,letterpaper]{article}
\usepackage{amsmath,amssymb,pgf,pgfarrows,pgfnodes,float,appendix, hyperref}
\usepackage{graphicx}
\usepackage{subfigure}

\usepackage[margin=0.9in]{geometry}
\usepackage{pgfplots}
\usepackage{amsmath}
\usepackage{tikz}

\title{{\rm\footnotesize \qquad \qquad \qquad \qquad \qquad \ \qquad \qquad \qquad \ \ \ \ \ \                  RUNHETC-2019-13 }\vskip.5in    Instantons, Colloids and Convergence of the 1/N Expansion for the Homogeneous Electron Gas }
\author{Tom Banks and Bingnan Zhang\\
Department of Physics and NHETC\\
Rutgers University, Piscataway, NJ 08854\\
E-mail: \href{mailto:bz173@scarletmail.rutgers.edu}{bz173@scarletmail.rutgers.edu}}

\date{}
\begin{document}
\maketitle

\begin{abstract}
We investigate non-perturbative corrections to the large $N$ expansion of the homogeneous electron gas. These are associated with instanton solutions to the effective action of the plasmon field.  We show that, although the large field behavior of that action dominates the quadratic bare Coulomb term, there are no solutions at large field, and consequently none at large density.  We argue that solutions would exist at low density if the large $N$ theory had a Wigner crystal (WC) phase.  However, we argue that this is not the case.  Together with the implied convergence of the large $N$ expansion, this implies that the homogeneous electron gas with $N$ component spins and a Coulomb interaction scaling like $1/N$ can only have a WC phase below a curve in the plane of $N$ and density, which asymptotes to zero density at infinite $N$.   We argue that for systems with a semi-classical expansion for order parameter dynamics, and a first order quantum transition between fluid and crystal phases, there are instantons associated with the decays of meta-stable fluid and crystal phases in the appropriate regions of the phase diagram.  We argue that the crystal will decay into one or more colloidal or bubble phases\cite{kivspiv} rather than directly into the fluid.  It is possible that the bubble phases remain stable all the way to the density where the crystal solution disappears.  The bubbles of fluid will expand as the density is raised and gradually convert the system into a fluid filled with chunks of crystal. The transition to a translationally invariant phase is likely to be second order. Unfortunately, the HEG does not have a crystal phase at large $N$, where these semi-classical ideas could be examined in detail.  We suggest that the evidence for negative dielectric function at intermediate densities for $N = 2$ is an indicator of this second order transition.  While the infinite $N$ limit does not have a negative dielectric function at any density, it is possible that the closed large $N$ equation for the plasmon two point function, derived in\cite{ergheg} might capture at least the qualitative features of the second order transition.

\end{abstract}

\section{Introduction}

In a recent paper\cite{ergheg}, one of the authors (TB) introduced a closed equation for the two point correlation function of the plasmon field in the homogeneous electron gas (HEG), as a large $N$ approximation to an exact equation for the derivative of the correlation function with respect to the dimensionless chemical potential.  In order to solve that equation for the correlation function, one needs a boundary condition at large chemical potential, where the leading large $N$ expressions appear to be adequate approximations to the exact result.

This paper began as an investigation of whether this appearance of adequacy could be misleading.  The expressions obtained for correlation functions to all orders in the $1/N$ expansion have expansions in terms of inverse powers of $\mu$ and powers of ln $\mu$.  Corrections of the form $e^{- \mu^p}$ with positive $p$ would of course be invisible in this expansion.  In relativistic field theories we know of two sources of this kind of non-perturbative ambiguity.  The first is ultraviolet divergence of coupling constant renormalization, also known as {\it renormalon} singularities in the literature.  The HEG has only a simple normal ordering divergence, so it has no renormalons. In addition, in relativistic field theories, renormalons do not affect the convergence of the large $N$ expansion. The other conventional sources of non-perturbative ambiguities are {\it instantons}, non-trivial stationary points of the imaginary time functional integral.  In the first part of this paper we will argue that for sufficiently large $\mu$, no finite action stationary points exist.

The absence of instantons for large $\mu$ also has implications for convergence of the $1/N$ expansion in this regime.  The $1/N$ expansion should converge for large enough $\mu$.  Thus, our assumption that the large $\mu$ behavior of the theory at large $N$ is accurately determined by the large $\mu$ behavior of terms in the $1/N$ expansion appears valid.   The solution of the closed equation for the two point function derived in \cite{ergheg} is thus completely determined and should give an accurate (non-perturbative in $N$) description of the theory for sufficiently large $N$.  

In \cite{dftfft1} TB argued that at large $N$ the HEG undergoes a first order phase transition to a WC phase.   The argument was a simple adaption of Wigner's classic argument, but it turns out to be {\it wrong}.  In fact we will argue that, in the Hartree approximation, which is exactly valid at large $N$, there is no WC.  Both terms in the Hartree functional are minimized by a constant density.  We also show that the functional is strictly convex, so that there is not even a meta-stable WC, and argue that this means there are no instantons.  This suggests strongly that the large $N$ expansion of the HEG has a finite radius of convergence for all $N$ independent density.  We will argue that recent numerical work by Haule {\it et. al.}\cite{haule} provides some evidence for this conjecture.  By directly examining the low density limit at finite $N$, we suggest that the transition density to the WC scales as an inverse power of $N$.

It might be possible to see the WC at large $N$, by introducing, for two dimensional systems, a constant magnetic field.  We will not engage in a detailed discussion of this system in this paper, but give only qualitative outlines.  The field introduces a new length scale, the Larmor radius, into the problem.  The kinetic energy of electrons has degenerate eigenstates corresponding to different guiding center solutions in the lowest Landau level (this is an overcomplete basis).  The expectation value of the Coulomb interaction is minimized for guiding centers that sit on a triangular lattice, whose spacing is determined by the density.  Note that there is no translation invariant superposition of guiding center states for any finite number of particles.  Thus, when the magnetic field is strong enough that we can neglect higher Landau levels, and the electron density is small enough that the Slater determinant of guiding center states centered at lattice sites is non-zero, we can expect the Wigner crystal will be the ground state.  Note that, because of the degeneracy of the Landau levels there is no longer any reason to prefer the wave function totally antisymmetric in $SU(N)$ "spin" and symmetric in space even at large $N$.  We will return to the large $N$ treatment of the HEG in a strong magnetic field in a future publication.  

Of course, in a large magnetic field, the HEG has a large number of topological Fractional Quantum Hall Phases.  These are all believed to be homogeneous gapped insulating states, which support a variety of topological field theories describing a discrete ground state degeneracy.  The system has no true fluid phase when only the lowest Landau level is occupied. Thus, our intuitive picture of the colloidal phase may also not describe the system in high magnetic field.  This is definitely a topic to which we will return.

\section{Instantons, Phases and Colloids}

In terms of rescaled variables\cite{ergheg} the effective action for the plasmon field is
\begin{equation} 
S = N \frac{\mu^{1/2}}{8} \int d^4 x\ \phi (- \nabla^2)\phi + {\rm tr\ ln}[\partial_t -\frac{\nabla^2}{2} - 1 + i \phi ]
\end{equation}
This is complex.  An instanton corresponds to a finite action stationary point of this action, which lies on a steepest descent contour into which the original integration contour can be deformed.  We will first investigate the existence of instantons at large $\mu$, in the fluid phase of the model.  Instanton solutions must approach the vacuum solution $\phi = 0$ at infinity in space and time.  The first term in the action has no finite action solutions with this property, so they could only exist for large $\mu$ if there is some region of field space where the determinant dominates the quadratic term.  
There is of course such a regime for fields with support only at small spatial momentum, but adding the quadratic term in the determinant, which gives Debye screening, does not lead to an instanton solution.  The remaining region that must be explored is the region of large $\phi$ .   The Schrodinger operator $(\partial_t - \nabla^2/2 - 1 + i g \phi (t,x)) $ is transformed into $g (\partial_s - \nabla_y^2/2 - 1/g + i \phi (s/g , y/\sqrt{g}))$ by rescaling the space and time variables, so going out to large $\phi$ in some direction in field space, is equivalent to going to slowly varying fields.

To compute the determinant of the Schrodinger operator, put the system on an imaginary time circle with anti-periodic boundary conditions.   The eigenvalue equation 
\begin{equation} (\partial_t + \nabla^2/2 + 1 + i \phi) \psi = \lambda \psi\end{equation}
can be formally solved by 
\begin{equation} \psi (\beta) = e^{\lambda \beta} U(\beta) \psi (0) ,\end{equation} where
\begin{equation} U(\beta ) = Te^{ - \int_0^{\beta} ds ( - \nabla^2 / 2 + 1 + i \phi (s,x) )} . \end{equation}
As we will recall in Appendix B, the time ordered product has an expansion, analogous to the Cambell-Baker-Hausdorff/Zassenhaus expansions, in terms of products of exponentials of multiple commutators of the time dependent Hamiltonian $ - \nabla^2 / 2 + 1 + i \phi (s,x)$.  The multiple commutators involve higher and higher spatial derivatives of $\phi$ and so will be small in the large $g$ limit. The leading term is obtained by just dropping the time ordering symbol, which shows that for large fields the action is just $\beta E$, where $E$ is the "energy" in a constant potential $i \bar{\phi}$, which is the time average of the putative instanton configuration.    Thus, the large $\phi$ limit of the equation for instantons is just the static Thomas-Fermi equation ($V = i \bar{\phi}$)
\begin{equation} \frac{\mu^{1/2}}{4} ( \nabla^2 V ) + \frac{5}{2}(V)^{3/2} = 0 . \end{equation}   The boundary condition is that $V$ must vanish at infinity.  If $V = a U$ with $ \frac{5}{2}(a)^{1/2} = 
 \frac{\mu^{1/2}}{4} $, then 
\begin{equation} ( \nabla^2 U ) + (U)^{3/2} = 0 . \end{equation}
By conventional arguments\cite{colemanetal} a spherically symmetric solution will have lowest action and will satisfy
\begin{equation} (\partial_r^2 + \frac{2}{r} \partial_r ) U + U^{3/2} = 0 . \end{equation}
This has an exact solution
\begin{equation} U_0 = 144 r^{-4} ,\end{equation} which has infinite action.  Writing $U = r^{-4} f$ and introducing $s = {\rm ln}\ r$ as the independent variable we obtain
\begin{equation} \partial_s^2 f - 6 \partial_s f + 12 f + f^{3/2} = 0 . \end{equation}
Now we can reduce the order of this equation by introducing $D(f) \equiv \partial_s f$.  The finite action boundary conditions imply that $f \rightarrow 144$ as $s \rightarrow \infty$ and $f \rightarrow 0$ as $s \rightarrow - \infty$.  The equation for $D(f)$ is
best written in terms of the independent variable $y = f^{1/2}$.  Then
\begin{equation} D (\partial_y D - 12y)+2 y^3(12+y) = 0 . \end{equation}  In order to have a finite contribution to the action near $r=0$, we want a solution with a power series expansion $ D = \sum_{n = 2}^{\infty} D_n y^n $.  In order to have a finite contribution from the region of large $r$, we must have $D ( - 12) = 0$.  

The iterative solution for the power series coefficients is
\begin{equation} D_2^2 - 6 D_2 + 12 = 0 . \end{equation}
\begin{equation} D_3 = \frac{-2}{5D_2-12} . \end{equation}
\begin{equation} D_{k} = (12 - (k + 2)D_2 )^{-1} \sum_{n = 3}^{k-1} (k+2-n) D_n D_{k + 2 - n} \ \ \ \ k\geq 4 . \end{equation}
We can extrapolate this series to arbitrary complex values of $f = y^2$ by Pade approximation, and it does not have a zero at $y = -12$.  Details can be found in Appendix C. Thus, there is no finite action solution of the spherically symmetric ansatz, which gives a lower bound on the action of non-spherical solutions.

A more general argument, which avoids the assumption of spherical symmetry follows from the work of Lieb and Simon\cite{liebsimon}.   They showed that the Thomas-Fermi equations, in an external potential had a unique lowest energy solution for each choice of potential.  For a periodic potential, the fermion density has the period of the potential. Using their argument for a periodic potential, whose period goes to infinity, we find only the homogeneous solution.

There is a physical argument that an instanton should exist if the Wigner crystal phase is stable for some value of $\mu$.  In\cite{dftfft1}, TB showed that, at large N, the fluid phase was locally stable for all $\mu$ and conjectured that there was a first order phase transition to a crystal phase at some value $\mu_c$.  The meta-stable fluid phase below $\mu_c$ would then decay via bubble formation into the crystal, as first described in\cite{JulesVerne}.  Assuming a crystal solution of the large $N$ field equations, with lower energy density than the homogeneous solution, it's easy to construct the bubble solution in the thin wall approximation.

However, we have realized that Wigner's argument for the existence of the crystal is incorrect at large $N$, and we will show that there is not even a meta-stable crystalline solution to the large $N$ field equations.   The argument parallels one of Lieb and Simon\cite{liebsimon} for the Thomas-Fermi approximation to these equations.  In\cite{dftfft1} TB showed that the problem of finding static solutions of the large $N$ field equations for the plasmon field was equivalent to finding minima of the Hartree approximation to the density functional
\begin{equation} F_{Hartee} [n(x)] = F_0 [n(x)] + \int d^3 x \frac{(n(x) - n_0)(n(y) - n_0)}{|{x - y}|} . \end{equation} Here $n_0$ is the uniform background charge density, defined so that $\int (n(x) - n_0) = 0$.  $F_0$ is the minimum expectation value of the electron kinetic energy, in states constrained to have expectation value of the electron density operator equal to $n(x)$.  

It is obvious that both terms of this expression are positive and that both vanish for a uniform $n(x)$ so the fluid phase is the stable ground state for all values of $n_0$.  If there were a meta-stable crystal, then, along any path in $n(x)$ space connecting the crystal configuration to the constant solution, the density functional would take on a maximum.  We can then try to minimize the value of $F_{Hartree}$ at the maximum, searching over all paths connecting the two stationary points.  The configuration with the lowest maximum will be a stationary point with a negative fluctuation mode.  However, we will show in a moment that both terms in $F_{Hartree}$ are strictly convex.   No stationary point can have a negative mode. 

The statement of convexity is obvious for the Coulomb term.  $F_0$ on the other hand is the Legendre transform of the ground state energy of non-interacting electrons in an external potential $V(x)$ .   Therefore 
\begin{equation} \frac{\delta^2 F}{\delta n(x) \delta n(y)} = - [\frac{\delta^2 E_0 [V]}{\delta V(x) \delta V(x)}]^{-1} , \end{equation} where the inverse is meant in the sense of integral operators.  The right hand side is a negative definite operator since the second order perturbative correction to the ground state energy is negative for any quantum system.  $F_0$ is therefore convex.  In Appendix A we work out explicitly what happens to a system whose Hartree wavefunction is a crystalline sum of Gaussians, with the width of the Gaussian taken as a variational parameter.  As expected from our general argument, the Coulomb energy is monotonically decreasing as the width of the Gaussian is taken larger than the putative lattice spacing.

We believe that the fact that the Hartree approximation to the HEG does not give a Wigner crystal is known to some in the community of researchers who work on this problem.  We have not been able to find an original reference but the literature is replete with statements to the effect that "naive approximations to the density functional do not give the Wigner crystal".  What is novel about our analysis is that it is formulated within a systematic approximation scheme.   Furthermore, as we've indicated above, it also implies that the large $N$ expansion of the HEG has a non-zero radius of convergence.   Thus we can conclude with some confidence that there is a finite range of $N$, for which the the transition density to the Wigner crystal goes like an inverse power of $N$ when the magnetic field is zero.   We have also found statements in the literature, claiming that the Hartree Fock approximation does give a Wigner crystal solution\cite{hfwc}.
Our results disagree with these claims, even when we include the exchange correction in Appendix A.  More precisely, we find that in a putative crystalline HF ansatz, the width of the single particle wave functions must be of order the Wigner lattice spacing, so that the single particle density expectation value does not show clear crystalline peaks.  The canonical arguments for the Wigner crystal suggest a lattice spacing to width ratio that scales like $r_S^{1/4}$.  The numerical results of \cite{hfwc} instead give a ratio of order $1$ independent of $r_S$, consistent with our results in Appendix A.

\subsection{The low density limit {\it or} Why large $N$ misses the WC }

Our analysis does not say anything definitive about the physical case of $N = 2$, and leaves open the possibility that the WC exists for all finite $N$, below a density that scales like an inverse power of $N$. This is not true in the Hartree-Fock approximation, but that approximation is actually not valid
in the low density limit.   Following Wigner, we treat that limit by fixing the number of electrons and expanding the multi-body potential around its minimum, which is a regular lattice whose spacing is of order $r_S$.  For $K$ electrons, the potential actually has $K!$ minima related by a permutation of which electron is put on which lattice point.  We must then decide on the spin configuration, which is of course dependent on the choice of $N$.  Consider, for simplicity, the completely polarized configuration, which exists for all $N$.  For a given choice of minimum, the Hamiltonian is
\begin{equation} H = \sum P_i^2  + n_0 (\Omega^2)_{ij} y_i y_j , \end{equation} where $y_i$ is the deviation of the $i$-th electron coordinate from the $\Pi (i)$-th lattice point ($\Pi$ is a permutation).  This equation is solved by going to normal modes and the ground wave function is a product of Gaussians in the normal mode coordinates.  In another publication we will detail our argument that, although the width of low frequency mode wave functions can be arbitrarily large, the spread of the $y_i$ around zero is of order $r_S^{3/4}$, while the lattice spacing of the crystal is of order $r_S$\footnote{However, we have found that for a one dimensional crystal, there is a logarithmic divergence in the coefficient of $r_S^{3/4}$, coming from low frequency oscillation modes.  Thus, we would claim that the one dimensional Wigner crystal does not exist at any $N$. These calculations and others relating to the extreme low density limit, will appear in a separate paper\cite{tbbzld}.}  Thus, in the limit of infinite $r_S$, the density distribution indeed has a crystalline form\footnote{This is of course the standard claim in the literature\cite{carretal}\cite{book}. We have not found a systematic derivation of it that treats antisymmetrization properly.}.   Note that in principle we have to antisymmetrize this wave function by summing over permutations of which fermion coordinate is assigned to which lattice site.  Although the individual terms in this sum have only two body correlations, the sum itself is not of the form of a symmetric product of two body terms multiplied by a Slater determinant, the form used in Quantum Monte Carlo (QMC) simulations.   Fortunately, for asymptotically large $r_S$ the interference terms between different summands in the full wave function are exponentially small.

Parenthetically, it should be pointed out that the ratio between the lattice spacing and the Gaussian width of the wave functions is not very large at the transition points to the Wigner crystal found in QMC calculations.  Indeed, we have not calculated the numerical coefficients of $r_S^{3/4}$ in the Gaussian widths, and the reliability of these low density forms of the wave function at finite $r_S$ is questionable.  As noted in the previous footnote, for the one dimensional system we find a logarithmic divergence in the width.  Anharmonic corrections to the potential will introduce complicated multi-body correlations, and the semi-classical expansion around the classical crystal is clearly asymptotic, since the harmonic approximation already gives us $e^{ - c r_S^{1/4}}$ corrections to all quantities.

An alternative approach to the low density limit is to work in Fock space.   We have not yet worked out the details of this analysis, but there is one clear lesson in it, which throws light on our large $N$ results.  Naively, the argument that the WC exists is that the Coulomb interaction dominates the kinetic term in the low density limit.  Writing the Coulomb interaction in terms of the electron number density operator $\rho = N^{-1} \Psi^{\dagger} \Psi$, it is
\begin{equation} \int dxdy\ (\rho (x) - n_0) (\rho (y) - n_0) |x - y|^{-1} , \end{equation} where $n_0$ is the background positive charge density.  This commutes with the density $\rho (x)$ and has the same form as the Hartree term.  The difference is that $\rho (x)$ has only discrete eigenvalues\footnote{More properly, $\int \rho$ over any small region has only discrete eigenvalues.} whereas the expectation value of the density, which appears in the Hartree energy, is continuous.  Thus, although the Hartree energy is minimized for uniform density, the Coulomb energy operator cannot be so minimized, and a lattice forms instead.   In the large $N$ approximation $\rho = N^{-1} \sum \Psi_i^{\dagger} \Psi_i $, and has eigenvalues quantized in units of $1/N$.  Thus, we would expect to find crystal formation only at a density that scales like $1/N$ .   This argument is far from rigorous, because the kinetic energy is a singular correction to the Coulomb interaction.  In first quantized formalism the singularity is dealt with by making the harmonic expansion of the potential around the crystal and solving the harmonic Schrodinger equation exactly.  We hope to return to a field theoretic treatment of the low density limit in a future publications\cite{tbbzld}. 

Wigner's argument for the WC uses intuition about classical point particles.  From the point of view of quantum field theory this is not at all the same as the limit of classical fields.  The large $N$ limit treats the plasmon field, and thus the density operator, as classical variables, which take continuous values.  These can always spread out to cancel the uniform background charge density.  In the extreme low density limit the charge density, averaged over large enough distances, is still classical in the sense that its commutator with its time derivative is small in the ground state, but it still has quantized values and cannot exactly cancel the background.  The WC state owes its existence to that quantized spectrum.  At large $N$ it will appear only at densities of order $1/N$.  This explains the discrepancy between our large $N$ results and {\it e.g.} those of \cite{QMC} and other Quantum Monte Carlo papers for $N = 2$.  Figure 1 is a sketch of our proposed phase diagram for the HEG as a function of $N$ and the density.

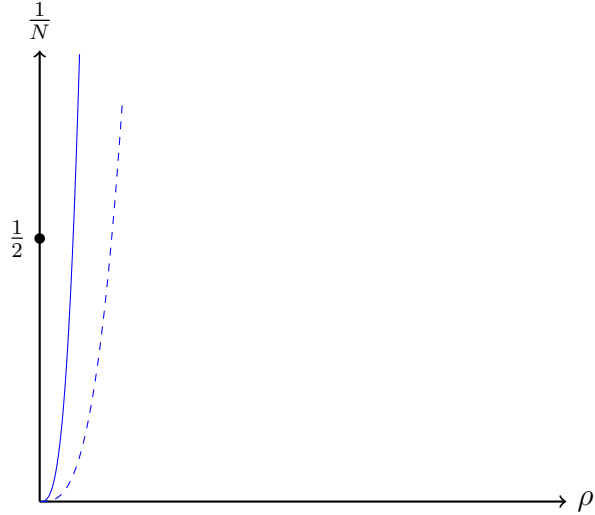
\begin{figure}[H]
\begin{center}
\begin{tikzpicture}
\draw [thick,->](0,0)--(7,0) node (xaxis)[right]{$\rho$};
\draw [thick,->](0,0)--(0,6) node (yaxis)[above]{$\frac{1}{N}$};
\fill [black](0,3.5)circle(2pt);
\node at (-0.3,3.5) {$\frac{1}{2}$};
\draw[scale=0.5,domain=0:2.2,smooth,dashed,variable=\x,blue] plot ({\x},{\x*\x*\x});
\draw[scale=0.5,domain=0:1.06,smooth,variable=\x,blue] plot ({\x},{10*\x*\x*\x});
\end{tikzpicture}
\caption{The Dashed Line is the Transition from Fluid to Colloidal Phases, the Solid Line is the Transition to the Wigner Crystal}
\end{center}
\end{figure}

\subsection{Summary}

We've argued that the large $N$ approximation to the HEG at fixed density has no meta-stable crystal phase and that the crystal is likely to appear only at densities that scale like $1/N$.  
For practical applications to real materials and experiments the existence of the zero field Wigner crystal does not seem terribly relevant.  The crystal structure of most materials has much higher electron densities than those at which the WC forms in the pure HEG.   For these applications, the fact that the $1/N$ expansion converges in the fluid phase is the most significant aspect of our work.  It provides a systematic, convergent calculation of the density functional in the regime of densities appropriate for at least some real materials.  The equation for the density density correlation function discovered in\cite{ergheg} resums an infinite number of diagrams in this expansion and should be more accurate than the simple $1/N$ expansion originally proposed in \cite{dftfft1}.  The next set of equations in the hierarchy proposed in \cite{ergheg} resums an infinite number of $1/N$ diagrams for the two, three and four point charge density correlators.  It is considerably more complicated than the two point equation, but the stakes are so high that it might be worth making the effort to solve it numerically.  

The result of this simple observation is quite dramatic.  It implies that there is no WC in the infinite $N$ limit.  When combined with our results about large field instantons this implies that there are no finite action instantons at all, so that we expect the $1/N$ expansion to have a finite radius of convergence.  This implies that there is a maximal $N$, above which the WC does not exist, even as a meta-stable state, for any $N$ independent density.  This is of course completely consistent with the existence of the WC for $N = 2$.  Indeed, we've given an argument that the WC should exist below a density of order $1/N$.

Finally, we remark on the significance of these results for the large $N$ truncations of the hierarchy of scaling equations for plasmon correlators.  Those equations relate the derivative of the one plasmon irreducible correlator $\Gamma_k$ w.r.t. the logarithm of the chemical potential, to integrals involving $\Gamma_l$ with $l \leq k +2$.  The large $N$ scaling of $\Gamma_l$ allows us to truncate this hierarchy by replacing $\Gamma_{k + 1}$ and $\Gamma_{k+2}$ by their large $N$ approximation.  In addition, the boundary condition at large $\mu$ is assumed to be read off the leading $N$ approximation.  We began this research in an attempt to establish that this method of imposing the large $\mu$ boundary condition was not missing important non-perturbative contributions.  We showed that at large $\mu$ the Thomas-Fermi approximation guaranteed the absence of instantons, so the boundary condition is precisely reliable.  

If the large $N$ expansion is convergent, then even the simplest truncation of the hierarchy, which gives a closed equation for $\Gamma_2$, whose solution resums an infinite number of large $N$ diagrams\footnote{More precisely the equation sums up all diagrams for the two point function containing only the leading and next to leading order (in N) vertices $\Gamma_{3,4}$, calculated using only one fermion loop. }, should give a reliable calculation of the density functional, down to some "reasonable" finite value of $N$.  One would expect this approximate density functional to be a good approximation to the exact result for $N = 2$ at all densities for which the paramagnetic fluid phase exists, and to give a reasonable approximation to the behavior near symmetry breaking second order phase transitions which terminate that phase.  It's unlikely that it gives a good description of the dynamics of symmetry breaking order parameters beyond those transitions.  If there is a ferromagnetic fluid phase, a spin density functional would capture the order parameter dynamics.

However, as we'll discuss below, the evidence seems to suggest that the first transition encountered as the density is lowered is to a colloidal phase that breaks translation symmetry.  We will present an intuitive discussion of that phase but we have not yet constructed a formal apparatus, analogous to density functional theory, which could provide us with a systematic large $N$ approximation to the colloid.

\subsection{2D Systems in a Magnetic Field}

In this section we want to indicate why we believe that the leading large $N$ approximation {\it might} give a stable Wigner crystal phase for a two dimensional system in a large enough field.  We will use an alternate presentation of the large $N$ equations in terms of the Hartree wave function.

The external field of course couples to the spin of the electrons.  For $N = 2$, a strong field projects out a single spin component.  To preserve a large $N$ limit, we take $N$ even and couple the field to an $SU(N)$ generator with equal numbers of positive and negative eigenvalues, with absolute value $1$.  Then half the spin components are projected out, but $M = N/2$ are left.  

At large $N$, the number density operator $N(x) = \frac{2}{N} \sum_{a = 1}^{N/2}  \Psi_a^{\dagger} \Psi_a $ is a classical variable in states containing $\ll N$ electrons. The Hartree-Fock approximation thus becomes exact.  Up to $1/N$ corrections we can replace $N(x) \rightarrow \langle N(x) \rangle \equiv n(x)$ in the Coulomb term in the Hamiltonian. 
In the $k$ electron Hilbert space with $k \leq N/$, a wave function 
\begin{equation} \psi_{a^1 \ldots a^k} (x_1 \ldots x_k) = \epsilon_{a^1 \ldots a^k} P(x_1 \ldots x_k) ,\end{equation} where $P$ is a {\it permanent} formed from $k$ guiding center states $\psi (x, x_j)$, in the lowest Landau levels compatible with the total density\footnote{We work in finite volume, so that the degeneracy of each Landau level is finite.}, for any choice of the guiding centers $x_j$ such that the wave functions are linearly independent, will minimize the expectation value of the kinetic term in the Hamiltonian.  

The density operator expectation value will be \begin{equation} n(x) = \int d^{3(k - 1)} y |P (x, y_1 \ldots y_{k - 1}|^2 . \end{equation} Unlike the calculations presented in Appendix A for the zero field problem, the Gaussian widths of the guiding center wave functions are fixed by the magnetic field.
Thus, when the density is low enough that one can choose the guiding centers to be further apart than the Larmor radius of the guiding center wave functions, the Hartree energy will be minimized by choosing the guiding centers to lie on the sites of the triangular lattice with spacing compatible with the density.  Wigner's original argument thus becomes rigorously correct in the large $N$ approximation.  It's clear that the transition between translation invariant and crystal phases will occur at a density such that $r_S$ is of order the Larmor radius.   Of course, we also expect to see a hierarchy of FQHE insulating states in this system, so a lot more detail is needed.

The remarks above are intended as an heuristic argument only.  We hope to return to a more rigorous and quantitative discussion of the 2D electron gas in a magnetic field in a future publication.  We will close this section by outlining the implications of a first order transition to a crystal phase, when the system of interest has a semiclassical limit.

\subsection{Instantons, Melting Transitions and Colloids}

Suppose there were a first order quantum phase transition between a fluid and a crystal phase of the HEG, perhaps in a background magnetic field.  The fluid phase would be stable for $\mu > \mu_c$ and meta-stable for smaller $\mu$, while the crystal phase would be stable for $\mu < \mu_c$ and meta-stable for $\mu_1 > \mu > \mu_c$.  We know that $\mu_1$ is finite since there are never crystal solutions of the high density equations.  Let us pretend, contrary to what was shown above, that this transition can be described in a systemative $1/N$ expansion.  In fact, what we present is a general semi-classical analysis of a quantum first order phase transition from a crystal phase.

The $1/N$ expansion is a semi-classical expansion for the $\phi$ field, so at large $N$ we would have two local static minima of the effective action for $\phi$, corresponding to the two phases.  The effective action is real for time independent configurations, so every path in field space between those two minima, would rise up to a maximum.   The minimal value of the maximum of the action, along all possible paths, defines a submanifold of field space $\phi^* (x;a)$, the {\it minimax manifold}. Here $a$ are coordinates on the submanifold.  For the gas phase the submanifold is a point.  For the crystal it corresponds to the independent phonon modes of zero momentum.  Small oscillations around the minimax submanifold have exactly one negative mode.

Instantons correspond to Euclidean paths  $\phi (x,t)$, which approach one of the meta-stable minima as $t\rightarrow \pm\infty$, which stationarize the Euclidean action\footnote{The action is complex and one is really searching for a steepest descent point in complex field space, along a contour with the appropriate asymptotics.}.  A finite action instanton will approach the meta-stable minimum at spatial infinity, for all time, and the region in field space near the instanton gives a finite contribution to correlation functions, of order $e^{-N}$, implying that the $1/N$ expansion is not convergent to the exact functional integral.  A finite action instanton has a field that approaches that of one of the meta-stable minima in all asymptotic directions in space and time.  

Let's first examine possible decays of the homogeneous phase, below $\mu_c$.  In the large $N$ expansion, there are no long range correlations in this phase, so that we expect the approach to asymptotic behavior to be exponential.  Therefore, there is a compact region in $t, x$ space outside of which the instanton field essentially vanishes.  Somewhere in that time interval the field hits the minimax configuration.  By time translation invariance we can call that point $t = 0$.  Let the size of the region where the minimax configuration is non-vanishing be $R$.  Non-trivial angular dependence adds a positive term to the action so we expect the minimax configuration to be roughly spherical.  The {\it thin wall approximation} to the minimax models it as a region of crystal of size $\sim R$ plus a transition region of size $\ll R$ to the $\phi = 0$ solution.  The energy of this configuration is
\begin{equation} E \sim - \epsilon R^3 + \sigma R^2 , \end{equation} where $\epsilon$ is the energy density difference between the meta-stable solution and the stable one, and $\sigma$ is a positive surface tension of the "bubble wall".  The minimax solution is the critical point $R_c^{-1}  = \frac{3 \epsilon}{2\sigma}$, of this energy, which is obviously unstable to increasing $R$.  The instanton is a field configuration that vanishes as $t \rightarrow \pm\infty$ and then opens up a small bubble, which expands in Euclidean time to the critical bubble at $t = t_0$, over an interval $[t_0 - T, t_0 + T]$.  The time $t_0$ and the position of the center of the bubble are collective coordinates of the solution.  For an infinite volume system we must sum over "dilute instanton gas" solutions which are well approximated by sums of single instanton solutions at space time points separated by distances larger than $T$ in the time direction and larger than $R_c$ in the space direction.  When $T E(R_c) \gg 1$ the real part of the instanton action is proportional to $T$, so the stationary point in the $T$ direction, which minimizes the real part of the instanton action is smaller than $E(R_c)^{-1}$. 

Physically, the dilute instanton gas represents a superposition of probability amplitudes  for critical bubbles to nucleate at different places and times. Small quantum fluctuations will cause these bubbles to expand.  For a meta-stable homogeneous fluid, the expanding bubbles collide and "swallow up" the fluid phase rather rapidly, so the superposition corresponds to different histories of how the meta-stable fluid decays.   The different histories might have crystal defects at the points where bubbles collide, so the actual quantum state is likely to be a decoherent superposition of crystals with different distributions of defects.

We can now see the new feature that arises if the meta-stable phase is crystalline and the stable phase is a fluid.  A critical bubble now occurs in some region of the crystal lattice. When that region is larger in all dimensions than the lattice spacing, it is again roughly spherical.  However, the would be expanding bubble now imposes a strain on the crystal\footnote{Here we are invoking continuum elasticity theory in the absence of a more microscopic analysis.}, proportional to the derivative of $E(R)$ with respect to $R$ and the elastic energy stabilizes the bubble at a value of $R$ slightly above $R_c$.

Since the crystal structure is a consequence of spontaneous symmetry breaking, the system contains phonons, which have not been taken into account in the analysis of the previous paragraph.  The phonons will release the strain and the elastic energy propagates out to infinity and dilutes.   This is correct for a single bubble, but not for a dilute gas.   For a collection of expanding bubbles of uniform density, the phonons are trapped in standing wave patterns and each decay history leads to a configuration that is inhomogeneous, with a distribution of fluid bubbles of various sizes inside the crystal determined by the positions and times of the critical bubble nucleation.  Note that quantum mechanics gives us only a probability distribution for the spatial structure of this inhomogeneous, {\it colloidal} phase.  Since it's likely that the bubbles are of macroscopic size in meta-stable colloidal configurations, the interference of the quantum amplitudes for different structures are probably negligibly small.

Note also that while a small crystalline region of a particular colloidal configuration might still, depending on its size and shape, decay by critical bubble nucleation, it will decay into a different colloidal configuration rather than into the homogeneous phase.  Thus the stable colloid state is actually a superposition of quantum states of different colloidal geometries and theory only predicts a probability distribution for the actual geometry.   In the semiclassical approximation, where all this analysis is valid, the time scale for transitions between different colloidal geometries is very long.  

As we dial up the chemical potential, making the crystalline phase less and less stable, we expect the ratio of volumes filled with fluid and crystal to change continuously, and eventually to reach a phase where we have a suspension of small crystals immersed in the fluid.  In principle, this phase could decay into the fluid by slow melting of the crystalline surfaces.  It would have a long lifetime simply because the surface to volume ratio of a large crystal is small.  However, another alternative is that the crystals have negative surface tension.  Then, just as in our calculation above of the critical bubble size, there is an equilibrium chunk of crystal, whose size depends on the ratio of surface tension and (positive) energy density of the crystal.  For negative surface tension and positive energy density, this equilibrium is {\it stable}.

In fact, for a crystal of electrons in a uniform positive background, it's clear that the surface tension is negative.  The charge of the bulk electron crystal is screened, and the surface energy is reduced by keeping the number of electrons fixed and increasing the surface area.  Thus we expect the {\it sol} phase of the electron colloid to be completely stable for a range of densities.

It is not clear that the transition between crystal dominated colloid (gel) and fluid dominated suspension (sol) is associated with a singularity of thermodynamic functions.  The long range properties of the two states are not easily distinguishable. In the conclusions, we will discuss evidence that there {\it is} a second order quantum phase transition between the colloidal phase and the translation invariant fluid ground state.

In order to actually realize these ideas as a systematic description of colloidal phases of the HEG, we'd have to be able to generalize the idea of large $N$ instantons into the regime of densities of order $1/N$.  This does not seem entirely implausible.  There are many known discrete lattice models where instanton effects can be studied in systems with discrete spectra.  This is done by introducing the discreteness into a continuous field description by using the Poisson summation formula.  Perhaps a variation on this trick will generalize to the HEG.

\section{Spontaneous Breakdown of $SU(N)$ ?}

Quantum Monte Carlo calculations suggest the existence of polarized fluid phases, in which the $SU(2)$ symmetry of the HEG is spontaneously broken at densities well above the WC transition.
Universality arguments suggest the possibility that these are second order transitions.  The question we want to address in this section is whether there are analogs of these polarized phases at large $N$.  They correspond to spontaneous breaking of $SU(N)$ to some subgroup $G$, and would result in Nambu-Goldstone excitations whose number is $N(N - 1) - {\rm dim} (G)$.  We conclude immediately that the dimension of $G$ must be of order $N^2$ .  Indeed the low temperature partition function of the HEG would be dominated by Goldstone excitations, and perhaps a gas of very low energy fermions, if the system were a Landau Fermi liquid.  The free energy is of order $N$ in the large $N$ expansion, so the number of independent low energy fields must scale at most like $N$.

The symmetry breaking order parameter is plausibly a fermion bilinear, and these transform in the adjoint representation plus a singlet of $SU(N)$.  The breaking pattern is thus $SU(N) \rightarrow SU(K_1) \times SU(K_p )\times U(1)^{k}$, where $\sum K_i + k = N - 1$.  The observation of the previous paragraph implies that exactly one of the $K_i$ is $N - r$, where $r$ is of order $1$ in the large $N$ limit.   Group the fermions into $N - r$ and $r$ and introduce a  $r \times r$ matrix source $J (x)$, which is time independent.  We want to compute the energy, at leading order in large $N$ as a function of a spatially constant $J(x)$.  The effective action for the plasmon field is
\begin{equation} S_{eff} = N \frac{\mu^{1/2}}{8} \int \phi ( - \nabla^2) \phi + (N - r){\rm Tr\ ln} [\partial_t - \nabla^2/2 - 1 + i (\phi - i\frac{\rm tr J}{N - r})] + {\rm Tr\ ln} [\partial_t - \nabla^2/2 - 1 + i (\phi - i J) ]. \end{equation} 
The large $N$ saddle point for $\phi$ is just $\phi = 0$, which means that to leading order in large $N$ we are just computing the energy of free electrons in the presence of $J$.  For small $J$, the linear term in the energy vanishes by symmetry, and the quadratic term is minimized at $J = 0$.  Thus, the $SU(N)$ symmetric phase is the stable ground state at large $N$.

\section{Conclusions}

From a practical point of view, our most important conclusion is the convergence of the $1/N$ expansion of the zero field HEG, in the regime of densities where the electron gas is a fluid.  This might be a regime applicable to the study of real materials.  Taken together with the approximate truncations of the hierarchy of identities for coupling derivatives of one plasmon irreducible correlators, which give partial resummations of the $1/N$ expansion of charge density correlators and the density functional, this result suggests that we have a powerful new tool for condensed matter computations.  

Light will be shed on this issue by extending our results to the case of two dimensional electron fluids in strong magnetic fields, where Wigner crystals have been observed in experiments.   In the text we gave an intuitive argument that the Wigner crystal is likely to survive in the infinite $N$ limit, when the cyclotron radius of the Landau orbits is small compared to the interparticle spacing.  However, recent calculations we have performed when the system is compactified on a sphere, seem to contradict this argument.  We will definitely return to further analysis of this problem in the near future.

Finally, our analysis of instantons describing the decay of a meta-stable crystalline phase of the HEG, suggests a general semi-classical theory of quantum melting.
Our basic claim is that meta-stable solids decay into colloids rather than homogeneous fluid phases.  In our picture the gel regime of a colloid is a (probably decoherent) quantum superposition of states with bubbles of fluid trapped inside a crystalline matrix, with the distribution of positions and sizes of bubbles calculated from a dilute instanton gas approximation. 
As a single parameter controlling the relative energies of the two phases is varied, the ratio of average fluid to solid densities changes.   Generically one might expect that this would lead to a continuous crossover, rather than a phase transition, between a gel and a sol state of the colloid.  The system would transform from a foam of fluid bubbles inside a solid matrix, into a sol of solid crystals bathed in fluid.  It would be extremely interesting to find a theoretical model of quantum crystalization in which the density order parameter could be treated in a systematic semi-classical expansion, and all of the pictorial language we have used could be realized in terms of explicit classical solutions.  We had hoped that the large $N$ HEG would be that model, but it has no crystalline phase for fixed density in the infinite $N$ limit. Our direct examination of the low density limit for finite $N$, suggests that what is missing at large $N$ is the discrete spectrum of the average number density operator $\frac{1}{N} \sum \psi^{\dagger}_i \psi_i $, and that the transition will occur at densities of order $1/N$.  We hope to find an improved version of the large $N$ expansion, which can explore this regime and become a testing ground for the semiclassical theory of quantum melting.

Because there is a change of symmetry between the fluid and colloid phases, we expect a transition, most likely second order, between them.  This should show up as a gapless mode at nonzero momentum in the plasmon correlation function, followed by an apparent region of instability, associated near the transition  with a negative quadratic term in the action for an order parameter that breaks translation and rotation symmetry.  An ancient analysis due to Landau\cite{landau}\footnote{TB would like to thank Eliezer Rabinovici for explaining Landau's argument to him.} suggests that this order parameter should preserve the rotation symmetries of the BCC lattice in three dimensions and the triangular lattice in two.  
In this connection, we would like to recall the evidence for a region with negative compressibility, associated with a negative static dielectric function, that has been reported in the literature above values of $r_S$ that are substantially lower than those where the WC transition occurs for $N = 2$\cite{negcompress}\cite{haule}\cite{gapless}. There is also evidence that the transition to this regime is associated with the appearance of a gapless plasmon mode, indicative of a second order quantum phase transition\cite{gapless}\cite{haule}.  These calculations are based on truncations of the electron/plasmon Schwinger-Dyson equations, which do not follow from any systematic expansion or from Monte Carlo resummations of the high density expansion.   The QMC calculations of the ground state energy also indicate negative compressibility in the low density regime\cite{book} and the review\cite{negcompress} argues that negative dielectric function is to be expected in the crystalline phase.  Indeed, a spectral representation of the charge density two point function, assuming only a stable translation invariant ground state, shows that the dielectric function is positive if such a ground state exists.  It is of course well known that the RPA dielectric function, the leading large $N$ result, is positive for all density, consistent with our finding that there is no Wigner crystal transition at infinite $N$.

Combining these observations with the large gap between the density at which negative dielectric function appears, and that at which the transition to the Wigner crystal phase occurs, one is led to the conclusion that for $N = 2$ the homogeneous electron gas has a non-translational invariant, but non-crystalline ground state, for some intermediate range of densities.  The colloid phase, first discussed by\cite{kivspiv}, and for which we have proposed a semiclassical model, would seem to be a prime candidate to play this role.  The transition between colloid and fluid should be second order.  

There is a nice intuitive picture of this transition, starting from the description of the sol state of the colloid as a suspension of finite electron crystals with negative surface tension.  Imagine that as the density is increased, the negative surface tension goes to zero.  The size of the equilibrium crystal shrinks and it becomes microscopic, at the same time as its energy goes to zero.  Our hypothesis is that in the small tension limit, this classical excitation of the fluid should be treated as a quantum quasi-particle, which is the gapless excitation seen\cite{gapless}\cite{haule} in approximate treatments of the fluid phase, a densities far above the transition to the Wigner crystal.  The crystalline nature of the excitation would explain why this gapless mode has non-zero wave number.

The closed equation for the plasmon two point function\cite{ergheg}, which is non-perturbative in $1/N$, is a promising tool for a more systematic study of the gapless mode and the properties of the phase transition, from the fluid side.  We hope to return to a numerical study of that equation in future work.

However, we do not see any obvious way that even the more complex equations in the hierarchy proposed in \cite{ergheg} could access the physics of the colloidal phase, for whose existence we have argued. These equations were derived under the assumption of a homogeneous ground state. The stable excitations in the sol part of the colloidal phase change from gapless microscopic quasi-particles to classical chunks of crystal as the density is decreased below the second order transition.  Once one introduces an external potential due to nuclei, the chunks would tend to cluster near large $Z$ centers, but might break if the charge of a chunk was greater than that of the nucleus.  The dynamics in this regime appears very complex, and it's hard to see how it can be captured by a density functional.   The existence of a colloidal phase opens up a new area of quantum field theory, for which we may not yet have adequate tools.

\section{Appendix A - Hartree-Fock Approximation for Large $r_S$}

Since the background field is a constant, the interaction energy between electrons and the background is fixed, irrelevant to the density distribution of electrons. So below we only consider the electron-electron interaction. The system is stabilized by putting it on a finite volume three torus and fixing the total number of electrons. For this appendix we abandon the field theoretic formalism of the text, and work with first quantized wave functions for a fixed number $K$ of electrons.

In some of the literature on the Wigner crystal\cite{hfwc} it is emphasized that despite the putative crystalline phase existing only in the strongly coupled regime, a self consistent Hartree-Fock approach should be valid at very low density.  If the system volume is $V$, then at fixed density the Coulomb potential has $K!$ minima, where each electron is put at one of the points of a space filling BCC lattice, whose spacing is approximately $r_S$, the Wigner-Seitz radius.  Of course there are also continuous translations of these minima by translations which are not in the space group of the lattice.  In the limit of infinite volume, the different continuous sectors decouple from each other, leaving behind a residue of soft phonon modes for each fixed choice of lattice positions on the torus.  The approximation that appears appropriate in the large $r_S$ limit is to expand the potential around its minimum and solve the coupled oscillator problem, which leads to the phonon spectrum of the crystal.  Then one writes down the ground state wave function for a fixed minimum, and sums over minima of the multi-body potential, with fermionic minus signs.  Although the phonon spectrum and wave functions for the normal mode coordinates are identical for each minimum, the formula for the normal mode coordinates in terms of individual electron coordinates is not.  We want to emphasize that the widths of the wave functions of the long wavelength normal mode coordinates are very broad, so that the Hartree-Fock ansatz with individual localized orbitals having a width given by the naive scaling of kinetic and potential energies is not obviously a good approximation to the true ground state of the system.  Note also that the exact wave function for the ground state, given this harmonic approximation is not of the Jastrow form of a single Slater determinant multiplied by a product of two body correlation factors.  Instead, for a fixed minimum, the Gaussian wave function does have only two body correlations, but anti-symmetrization is performed on these correlated wave functions, rather than on a wave function for independent particles, which is then multiplied by symmetrized two body correlations.  

Nonetheless, in this appendix, we calculate the best Hartree Fock wave function for the putative WC, with Gaussian orbitals around each point.  Imagine two electrons oscillating around two points P and Q. We model their wave functions, and thus their probability distributions, as Gaussian.
\begin{equation}
\rho_1(\vec{x})=\frac{1}{(2\pi)^{3/2}\sigma^3}e^{-\frac{(\vec{x}-\vec{P})^2}{2\sigma^2}}\;\;\rho_2(\vec{y})=\frac{1}{(2\pi)^{3/2}\sigma^3}e^{-\frac{(\vec{y}-\vec{Q})^2}{2\sigma^2}}
\end{equation}
\par 
The Coulomb energy between them is
\begin{equation}
E_c=\frac{1}{(2\pi)^3 \sigma^6}\int d\vec{x} d\vec{y} \frac{e^{-\frac{(\vec{x}-\vec{P})^2}{2\sigma^2}}e^{-\frac{(\vec{y}-\vec{Q})^2}{2\sigma^2}}}{|\vec{x}-\vec{y}|}
\end{equation}
shift variable $\vec{x}\rightarrow\vec{x}+\vec{P}\;\vec{y}\rightarrow\vec{y}+\vec{Q}$
\begin{equation}
E_c=\frac{1}{(2\pi)^3 \sigma^6}\int d\vec{x} d\vec{y} \frac{
e^{-\frac{\vec{x}^2+\vec{y}^2}{2\sigma^2}}}
{|\vec{x}-\vec{y}-\vec{L}|}
\end{equation}
where $\vec{L}=\vec{Q}-\vec{P}$\\
Change variables $\vec{u}=\frac{\vec{x}+\vec{y}}{\sqrt{2}\sigma}\;\;\vec{v}=\frac{\vec{x}-\vec{y}}{\sqrt{2}\sigma}$
\begin{equation}
\begin{aligned}
E_c &=\frac{1}{(2\pi)^3}\int d\vec{u} d\vec{v} \frac{
e^{-\frac{\vec{u}^2+\vec{v}^2}{2}}}
{|\sqrt{2}\sigma\vec{v}-\vec{L}|}\\
&=\frac{1}{\sqrt{2\pi}}\int_0^{+\infty}dv\int_{-1}^{+1}d\cos{\theta}\frac{v^2e^{-v^2/2}}{\sqrt{2\sigma^2v^2-2\sqrt{2}\sigma v L \cos{\theta}+L^2}}\\
&=\int_0^{L/\sqrt{2}\sigma}dv...+\int_{L/\sqrt{2}\sigma}^{\infty}dv...\\
&=I+II
\end{aligned}
\end{equation}
where we have divided the integral into two regions.Now we expand the denominator with Legendre  Polynomials, and use the fact that $\int_{-1}^{+1}d\cos{\theta}P_n(\cos{\theta})=0$ for $n>0$, we have
\begin{equation}
\begin{aligned}
I &=\frac{1}{\sqrt{2\pi}}\int_0^{L/\sqrt{2}\sigma}dv\;v^2e^{-v^2/2}\frac{1}{L}\sum_{n=0}^{\infty}\int_{-1}^{+1}d\cos{\theta}\;P_n(\cos{\theta})(\frac{\sqrt{2}\sigma v}{L})^n\\
&=\frac{1}{L}\sqrt{\frac{2}{\pi}}\int_0^{L/\sqrt{2}\sigma}dv\;v^2e^{-v^2/2}
\end{aligned}
\end{equation}
\begin{equation}
\begin{aligned}
II &=\frac{1}{\sqrt{2\pi}}\int_{L/\sqrt{2}\sigma}^{\infty}dv\;v^2e^{-v^2/2}\frac{1}{\sqrt{2}\sigma v}\sum_{n=0}^{\infty}\int_{-1}^1d\cos{\theta}P_n(\cos{\theta})(\frac{L}{\sqrt{2}\sigma v})^n\\
&=\frac{1}{\sqrt{\pi}\sigma}\int_{L/\sqrt{2}\sigma}^{\infty}dv\;ve^{-v^2/2}=\frac{1}{\sqrt{\pi}\sigma}e^{-L^2/4\sigma^2}
\end{aligned}
\end{equation}
denoting $\frac{L}{\sqrt{2}\sigma}$ as t, we finally have
\begin{equation}
E_c=I+II=\frac{1}{L}\sqrt{\frac{2}{\pi}}(\int_0^tdv\;v^2e^{-v^2/2}+te^{-t^2/2})
\end{equation}
\begin{equation}
\frac{\partial E_c}{\partial t}=\frac{1}{L}\sqrt{\frac{2}{\pi}}e^{-t^2/2}>0
\end{equation}
So as $\sigma$ becomes larger, t becomes smaller, $E_c$ also becomes smaller. According to the uncertainty principle, the electron's kinetic energy also becomes smaller. To see it explicitly, we consider one electron with Gaussian wave function $\phi(\vec{x})=\frac{1}{(2\pi)^{3/4}\sigma^{3/2}}e^{-x^2/4\sigma^2}$
\begin{equation}
E_k=-\frac{1}{2}\int d\vec{x}\;\phi \nabla^2 \phi=\frac{1}{8(2\pi)^{3/2}\sigma^7}\int d\vec{x}\; x^2 e^{-x^2/2\sigma^2}\propto \frac{1}{\sigma^2}
\end{equation}
So the total energy of the system becomes smaller as we increase $\sigma$.
This result agrees with our large $N$ analysis.  If the Gaussian width is, as we've assumed above, much smaller than the lattice spacing, then the exchange contribution to the HF energy is negligible and the HF approximation loses its Russian.  The Hartree approximation becomes exact for all densities at large $N$ and we've already argued that the Hartree functional has no periodic minimum. The computation above is an explicit demonstration of this for a particular ansatz.

Rather than trying to complete the full HF calculation for widths of order the lattice spacing, we simply compute the expectation value of the density operator in a Gaussian HF state.  When the width is much smaller than the spacing, this shows a clear crystalline form, but this disappears as the width approaches the lattice spacing.  Thus, the HF calculation does not give us a crystalline ground state.  

\section{Appendix B - Calculation of Determinants for Slowly Varying Fields}

The operator $\partial_t + h(t)$ where $h(t)$ is a time dependent single particle Hamiltonian, has eigenstates satisfying
\begin{equation} (\partial_t + h(t) ) \psi (t) = \lambda \psi (t) . \end{equation}  $\psi$ is a vector in the Hilbert space in which $h(t)$ acts.  The solution of this equation is
\begin{equation} 
\psi (t) = T e^{ - \int_{-\beta/2}^t h(s) ds} e^{\lambda (t + \beta)} \psi (0) . 
\end{equation} 
Anti - periodicity implies that $\psi (0)$ is an eigenstate of $T(\beta) \equiv T e^{ - \int_{-\beta/2}^{\beta/2} h(s) ds} $, with eigenvalue $S (\beta)$  and that $\beta \lambda = - {\rm ln} (S) + \pi i (2n + 1)$   If we write
\begin{equation} 
T(\beta ) = e^{ -\beta\bar{H}}
\end{equation}
then the {\it average Hamiltonian} has a nested commutator expansion 
\begin{equation} 
\bar{H} = \beta^{-1}\int_{-\beta/2}^{\beta/2} ds\ h(s) + \frac{1}{2} \int ds dt\ \theta (t - s) [h(s) , h(t) ] + \ldots
\end{equation} 

For the HEG, we have \begin{equation} [h(s), h(t)] = [\nabla^2 , V(s) - V(t)] , \end{equation} so this terms is small when $V(s, {\bf x})$ is a slowly varying function of the spatial coordinate.  In the text, we used these formulae to calculate the fermion determinant for large potentials, which, by the scaling argument, are equivalent to slowly varying potentials.

\section{Appendix C - Pade Extrapolation of the Thomas Fermi Series}
Let's start from the equation 
\begin{equation}
\partial_r^2 U+\frac{2}{r}\partial_r U+U^{3/2}=0
\end{equation}. Suppose the large r expansion of $U$ is $U=cr^a+o(r^a)$, we get the equation 
$ca(a+1)r^{a-2}+c^{3/2}r^{3a/2}=0$, where "=" holds at leading order. If $\frac{3a}{2}> a-2$, we should have $a(a+1)=0$, $a=0$ or -1. Neither of them has finite action. If $\frac{3a}{2}=a-2$, we have $a=-4, c=144$.  If $\frac{3a}{2}>a-2$, we should have $c=0$, which is inconsistent. So the only possible large-r behavior of U is $U=\frac{144}{r^4}+o(r^{-4})$. So $f\rightarrow 144$ and $y\rightarrow -12$ as $r\rightarrow \infty$. From the definition $D=\partial_s f=r\partial_r f$, we see that $D\rightarrow 0$.

Now look at the transformed equation
\begin{equation}
D(\partial_y D-12y)+2y^3(12+y)=0
\end{equation}
Suppose it has a solution, we approximate it by the Pade form
\begin{equation}
D(y)=\frac{\sum_{n=2}^{N}p_ny^n}{\sum_{m=0}^{N}q_my^m}
\end{equation}
Note that $D(y)$ starts from $y^2$ at small y limit, because $D=\partial_s f=r\partial_r f=O(f)=O(y^2)$.
Substitute this into the above equation, we can solve for the coefficients $p_n, q_m$. There are always two sets of solutions related by a complex conjugation. Below is a plot of the case $N=7$. The behavior of $Re[D]$ and $Im[D]$ within the plotting range has already stabilized.
\begin{figure}[H]
\centering
\includegraphics[scale=0.6]{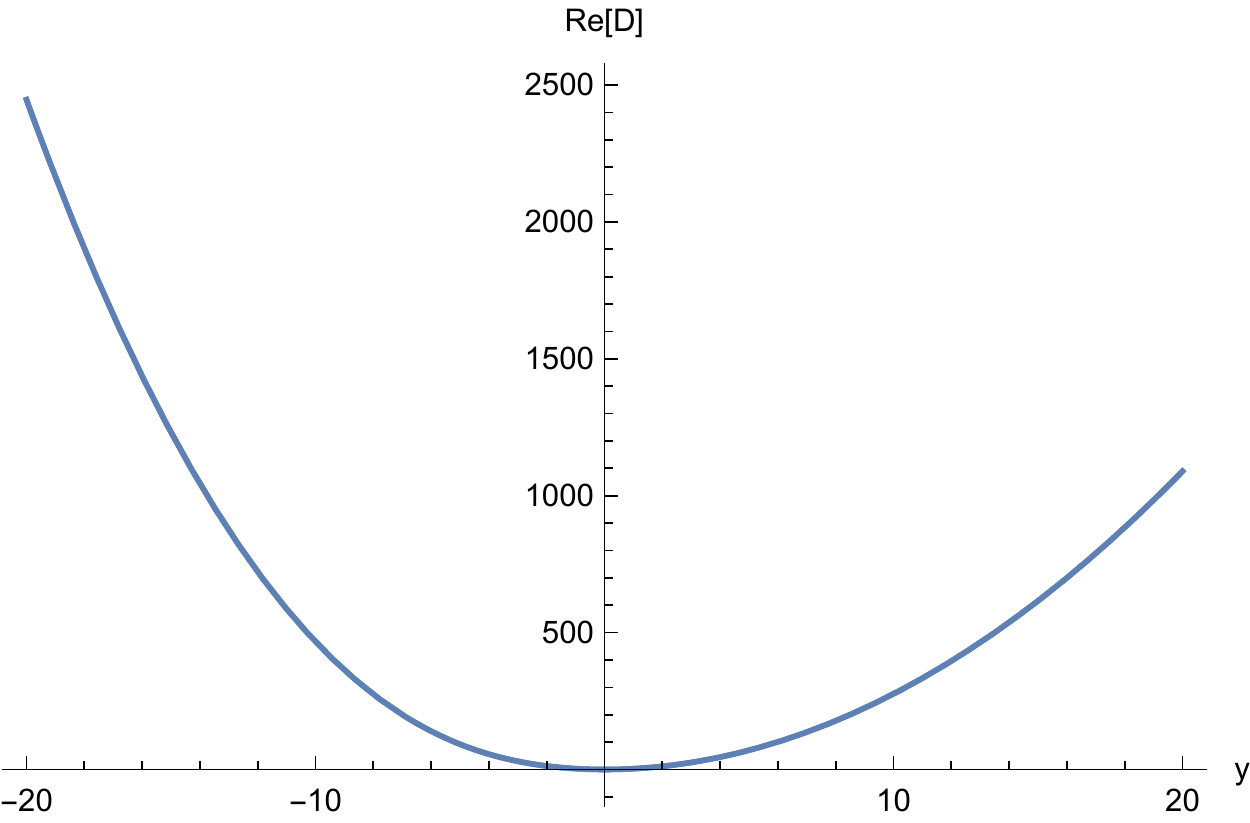}
\includegraphics[scale=0.6]{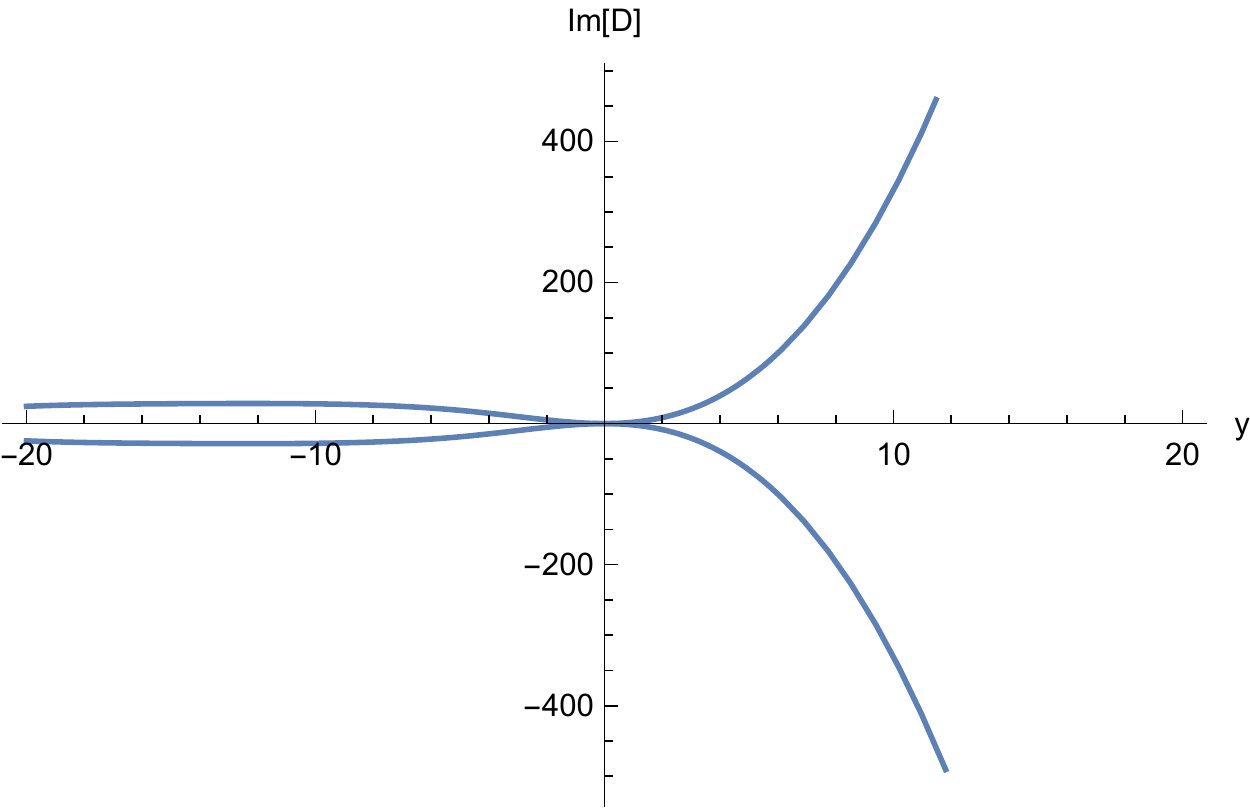}
\caption{Real and imaginary part of D(y)}
\end{figure}

 We found that the only stable zero point of $D(y)$ is at $y=0$, which shows that D cannot be zero at $y=-12$, which is the criterion for a finite action spherical solution.\\\\\\

\begin{center} 
{\bf Acknowledgments }  \end{center}

The work of T.Banks is {\bf\it NOT} supported by the Department of Energy, the NSF, the Simons Foundation, the Templeton Foundation or FQXi. TB would like to thank E.Rabinovici, S.Sachdev, D. Vanderbilt, E. Lieb, P. Gori Giorgi, K. Burke and especially K.Haule for crucial insights and helpful suggestions, and the entire condensed matter group at Rutgers university for patiently listening to an interloper on their territory and making numerous helpful comments.


\begin{thebibliography}{99}
\bibitem{kivspiv} S.A. Kivelson, B. Spivak, {\it Transport in two dimensional electronic micro-emulsions}
	Ann. Phys., 321, Issue 9, p. 2071-2115 (2006).
\bibitem{ergheg} T. Banks, {\it Broken Scale Invariance Ward Identities for the Homogeneous Electron Gas}, arXiv:1806.01749 [cond-mat.-st. el.] .
\bibitem{dftfft1}  T. Banks, {\it Density Functional Theory for Field Theorists I}, arXiv:1503.02925 [cond-mat.mtrl-sci] .
\bibitem{haule} K. Chen, K. Haule, {\it Feynman's Solution of the Quintessential Problem in Condensed Matter Physics},  arXiv:1809.04651v1  [cond-mat.mtrl-sci]  12 Sep 2018, Fig. 4d for evidence of gapless mode.
  \bibitem{liebsimon} E.H. Lieb, B. Simon, {\it The Thomas-Fermi Theory of Atoms Molecules and Solids}, Adv. Math 23, 22-116 (1977).
\bibitem{JulesVerne} J. Verne, {\it Off on a Comet}, Pierre-Jules Hetzel Publishing, 1877.
\bibitem{hfwc}  J.R. Trail, M. D. Towler, and R. J. Needs. "Unrestricted hartree-fock theory of wigner crystals." Physical Review B68.4 (2003): 045107.
\bibitem{landau}  L.D.Landau, Phys. Z. Soviet II26, 545 (1937).
\bibitem{tbbzld} T. Banks, B. Zhang, {\it Low Density Limit of the Homogeneous Electron Gas}, work in progress.
  \bibitem{carretal} W.J. Carr, {\it Energy, Specific Heat, and Magnetic Properties of the Low Density Electron Gas}, Phys. Rev. 122,5 (1961), 1437.
  \bibitem{book} G.F. Giuliani, G. Vignale, {\it Quantum Theory of the Electron Liquid}, Cambridge University Press, Cambridge, UK, (2005), pp. 59-62.
  \bibitem{negcompress}  O. V. Dolgov, D. A. Kirzhnits, and E. G. Maksimov,Rev. Mod.Phys.53,81 (1981).
  \bibitem{gapless} Y. Takada, {\it Excitonic Collective Mode and Negative Compressibility in Electron Liquids}, Journal of Superconductivity, Incorporating Novel Magnetism,18, Nos. 5/6, (2005).
\bibitem{colemanetal}  S.R. Coleman, V. Glaser, A. Martin, {\it Action Minima Among Solutions to a Class of Euclidean Scalar Field Equations} , Commun.Math.Phys. 58 (1978) 211-221. 
\bibitem{QMC} R.M. Martin, L. Reining, D.M. Ceperly, {\it Interacting Electrons: Theory and Computational Approaches}, (particularly Part IV), Cambridge University Press, (2016); D.H. E. Dubin, T. M. O'Neil, {\it Trapped nonneutral plasmas, liquids, and crystals (the thermal equilibrium states)}, Rev. Mod. Phys. 71, 1 87-172 (1999) and other QMC papers cited therein.
   \end{thebibliography}
\end{document}